\documentclass[twocolumn,showpacs,preprintnumbers,amsmath,amssymb,prx]{revtex4-1}
\usepackage{mathrsfs}
\usepackage{graphicx}
\usepackage{dcolumn}
\usepackage{bm}
\usepackage{amsmath}
\usepackage{amsfonts}

\begin{document}

\title{Quantum anomalous Hall effect in magnetic topological insulators}
\author{Jing Wang}
\author{Biao Lian}
\author{Shou-Cheng Zhang}
\affiliation{Department of Physics, McCullough Building, Stanford University, Stanford, California 94305-4045, USA}

\begin{abstract}
The search for topologically non-trivial states of matter has become an important goal for condensed matter physics. Here, we give a theoretical introduction to the quantum anomalous Hall (QAH) effect based on magnetic topological insulators in two-dimension (2D) and three-dimension (3D). In 2D topological insulators, magnetic order breaks the symmetry between the counter-propagating helical edge states, and as a result, the quantum spin Hall effect can evolve into the QAH effect. In 3D, magnetic order opens up a gap for the topological surface states, and chiral edge state has been predicted to exist on the magnetic domain walls. We present the phase diagram in thin films of a magnetic TI and review the basic mechanism of ferromagnetic order in magnetically doped topological insulators. We also review the recent experimental observation of the QAH effect. We discuss more recent theoretical work on the coexistence of the helical and chiral edge states, multi-channel chiral edge states, the theory of the plateau transition, and the thickness dependence in the QAH effect.
\end{abstract}

\date{\today}


\maketitle

\section{Introduction}
\label{sec1}

The search for topological states of quantum matter has attracted intensive interest in condensed matter physics~\cite{qi2011,hasan2010}.
The quantum Hall effect (QHE), discovered in 2D electron systems in 1980s~\cite{klitzing1980}, was the first
topological quantum state different from any other quantum states known before. In the QHE, the electronic states of the 2D electron system form Landau levels (LLs) under strong external magnetic fields, which is topologically distinct from vacuum.
The 2D bulk of the sample is an insulator, while the edge of the sample is electrically conductive. The flow of
this chiral current is dissipationless and results in a quantized Hall resistance $h/\nu e^2$, where $h$ is the Planck constant,
$e$ is the electron charge, and $\nu$ is an integer. The exactly quantized Hall resistance is because it is characterized by a
topological invariant $\nu$, which is the first Chern number~\cite{laughlin1981,thouless1982} and is independent of material details.

In principle, the QHE can exist without the external magnetic field and the associated LLs; however,
the Haldane model~\cite{haldane1988} with circulating currents on a honeycomb lattice is not easy to implement experimentally. The QAH effect, considered as a quantized version of anomalous Hall effect discovered in 1881~\cite{hall1881}, has been theoretically
proposed for magnetic topological insulators (TIs)~\cite{qi2006,qi2008,liu2008,li2010,yu2010,wang2013a}, where the ferromagnetic ordering and spin-orbit coupling (SOC) are sufficiently strong that they can give rise to a topologically nontrivial phase with a finite Chern number. The rich material choice of time-reversal-invariant TIs open up a new way to the experimental realization of the QAH effect. A 2D TI is expected to show the quantum spin Hall (QSH) effect~\cite{bernevig2006c}, in which, a pair of spin-filtered counter-propagating helical edge states flows without dissipation and contributes to a quantized longitudinal resistance. The QSH effect has been realized and observed experimentally in both HgTe/CdTe~\cite{koenig2007} and InAs/GaSb~\cite{liu2008a,knez2011} quantum wells. 3D TIs with a insulating bulk and gapless 2D Dirac-type surface states on each surface, have been observed in Bi$_x$Sb$_{1-x}$ alloy~\cite{fu2007,hsieh2008} and in Bi$_2$Se$_3$ family compounds. Bi$_2$Se$_3$, Bi$_2$Te$_3$, and Sb$_2$Te$_3$ compounds have a large bulk gap and single Dirac surface band~\cite{zhang2009,xia2009,chen2009}.

The QAH effect can be naturally achieved by introducing ferromagnetism (FM) into the TIs that breaks the time-reversal (TR) symmetry. In a 2D TI, FM can suppress one of spin channels of the QSH edge states, driving the system into a QAH phase~\cite{liu2008}. Magnetization in a 3D TI gaps out the Dirac fermions at each surface perpendicular to the magnetization vector, and leads to a QHE with a half QH conductance ($e^2/2h$)~\cite{qi2008}. In a 3D TI film with perpendicular magnetization, the gapped surface bands at top and bottom surfaces have distinct topological characters due to their opposite normal directions. Thus the edge of the sample can be viewed as a domain wall between the upper and lower surface bands, which induces a chiral edge state that carries a Hall conductance of $e^2/h$, giving rise to QAH effect. By tuning the Fermi level of the sample into the magnetically induced gap, one should observe a quantized Hall conductance plateau $\sigma_{xy}=e^2/h$ and a vanishing longitudinal conductance $\sigma_{xx}$ without a magnetic field.

In this paper, we give a theoretical introduction to the QAH effect based on magnetic TIs.
The organization of this paper is as follows. After this
introductory section, Section~\ref{sec2} briefly reviews the basic mechanism of the QAH effect in 2D and 3D magnetic TIs.
Section~\ref{sec3} describes the phase diagram in thin films of a magnetic TI. Section~\ref{sec4} presents the mechanism of
ferromagnetic order in magnetically doped TIs. Section~\ref{sec5} introduces the experimental observation of the QAH effect.
Section~\ref{sec6} presents the recent theoretical developments of the QAH effect. Section~\ref{sec7} reviews the recent
experimental progress. Section~\ref{sec8} concludes this paper.

\section{Basic mechanism of the QAH effect}
\label{sec2}

As discussed, breaking the TR symmetry of TIs through introducing FM can lead to the QAH effect. Here in this section,
we review the basic mechanism of the QAH effect in magnetic TIs.

\subsection{Spin polarized band inversion in 2D}
As a starting point, we cam consider a generic two-band Hamiltonian:
\begin{equation}\label{2bandmodel}
h(\mathbf{k}) = \epsilon(\mathbf{k})1_{2\times2}+d_a(\mathbf{k})\sigma^a,
\end{equation}
where $1_{2\times2}$ is the $2\times2$ identity matrix, $\sigma^a$ ($a=x,y,z$) are Pauli matrices, and
\begin{eqnarray}
\epsilon(\mathbf{k})&=&C-D(k_x^2+k_y^2),
\\
d_a(\mathbf{k})&=&[Ak_x,-Ak_y,M(\mathbf{k})],
\\
M(\mathbf{k})&=&M-B(k_x^2+k_y^2)
\end{eqnarray}
This two band model alone describes a TR symmetry-breaking system \cite{qi2006}. the system is an insulator with a quantized Hall conductance provided there is a gap between the two bands \cite{thouless1982}. The value of the Hall conductance is given by the first Chern number defined in the Brillouin Zone, which can be written as
\begin{equation}
\sigma_{xy} = \frac{e^2}{h}\frac{1}{4\pi}\int dk_x\int dk_y \hat{\mathbf{d}}\cdot\left(\frac{\partial\hat{\mathbf{d}}}{\partial k_x}\times\frac{\partial\hat{\mathbf{d}}}{\partial k_y}\right)
\end{equation}
for the generic two band model in Eq. (\ref{2bandmodel}). It has the topological meaning $e^2/h$ times the winding number of the skyrmion pattern of the unit vector $\hat{\mathbf{d}}(\mathbf{k})$
around the unit sphere. For $M/B>0$ and $M/B<0$, the winding number equals to 1 and 0, respectively. The system behaves like an ordinary QH insulator, with a chiral edge state that contributes to the Hall conductance $e^2/h$.
Such a model describes the QAH effect realized
with both strong spin-orbit coupling ($\sigma_x$ and $\sigma_y$ terms) and ferromagnetic polarization ($\sigma_z$ term).
The basic mechanism here is to realize the spin polarized band inversion ($M/B>0$) in 2D.

To realize spin polarized band inversion in 2D, one can break TR symmetry in the QSH system, where the bands are already inverted.
The QSH system is described by
\begin{equation}\label{qsh}
H_{\mathrm{QSH}} =\left(\begin{array}{cc}
h(\mathbf{k}) & 0\\
0 & h^*(-\mathbf{k})
\end{array}\right),
\end{equation}
due to TR symmetry, the upper and lower $2\times 2$ blocks carry the opposite winding numbers, and thus the edge states are helical and the total Hall conductance vanishes. A QSH insulator can therefore be understood as two QAHs which are TR partners. When the TR symmetry is broken, the helical edge states no longer propagate symmetrically, and the Hall conductance will become nonzero. In particular, if we consider the two blocks to have different masses $M$ (which breaks the TR symmetry) so that one block has $M/B<0$ and the other block has $M/B>0$, the whole system will become a QAH state with Hall conductance quantized to $e^2/h$. In the experiment, this can be realized by introducing an exchange coupling with magnetic impurities. Generically, the spin splitting term induced by the magnetization of magnetic impurities takes the form
\begin{equation}
H_s =\left(\begin{array}{cccc}
G_1 & 0 & 0 & 0\\
0 & G_2 & 0 & 0\\
0 & 0 & -G_1 & 0\\
0 & 0 & 0 & -G_2
\end{array}\right)
\end{equation}
where $G_1$ and $G_2$ are the splitting of bands 1 and 2,
respectively. This term effectively changes the mass term $M$ of the upper block to $M+(G_1-G_2)/2$, and that of the lower block to $M-(G_1-G_2)/2$. Provided $G_1\neq G_2$, the masses of the two blocks are different from each other, and the realization of QAH is possible. In the model considered here, the QAH can be realized if $G_1G_2<0$, while the system becomes metallic before developing a quantized Hall conductance if $G_1G_2>0$. The above discussion of TR breaking can be easily generalized to more realistic models, which serves as a guiding principle for generating a QAH insulator from a QSH insulator.

\subsection{Magnetic domain wall on 3D TI surfaces}
The low energy effective Hamiltonian of surface states of a 3D TI with a single Dirac cone is
\begin{equation}\label{surfaceH}
H_{\mathrm{surf}} (k_x,k_y) = v_F(\sigma^xk_y-\sigma^yk_x),
\end{equation}
where the $z$ direction is perpendicular to the surface and $v_F$ is the Fermi velocity. Now we consider the FM proximity to the surface state, and the perturbation Hamiltonian is $H_1=\sum_{a=x,y,z}m_a\sigma^a$. The total Hamiltonian has the spectrum
$E_{\mathbf{k}}=\pm\sqrt{(v_Fk_y+m_x)^2+(v_Fk_x-m_y)^2+m_z^2}$. Thus, only $m_z\sigma^z$ term can open a gap and destabilize the
surface states, which is odd under TR. This mass term will induce a half-integer QH conductance, which can be obtained by
Eq.~(5)
\begin{equation}\label{halfqh}
\sigma_{xy}=\frac{m_z}{|m_z|}\frac{e^2}{2h},
\end{equation}
here the $\mathbf{d}(\mathbf{k})=(v_Fk_y,-v_Fk_x,m_z)$ vector has a meron configuration and $\hat{\mathbf{d}}(\mathbf{k})$ covers half of the unit sphere.

This half-integer quantum Hall is unique to the 3D TI surfaces, originating from the nontrivial bulk topology \cite{qi2008}. However, unlike the usual integer quantum Hall effect, the half-integer quantum Hall on the TI surface is not a measurable effect. This is because in real systems the quantized Hall conductance always comes from the edge states, but mathematically the surface of a 3D TI is always a closed two-dimensional manifold that has no edge. Even if the entire TI surface is gapped out by magnetic impurities, the system cannot carry a charge Hall current due to the lack of an edge. On the other hand, if the TI is magnetically dopped and forms a ferromagnetic phase, there will be a domain wall of magnetization on the side surface, and thus the Hall conductance will jump by $e^2/h$ across the domain wall according to Eq. (\ref{halfqh}). Correspondingly, there will be a chiral edge state at the domain wall [see Fig.~1(a)]. This mechanism therefore provides us another way of realizing the QAH effect at zero magnetic field.

\begin{figure}[t]
\label{fig2}
\begin{center}
\includegraphics[width=3.3in,angle=0]{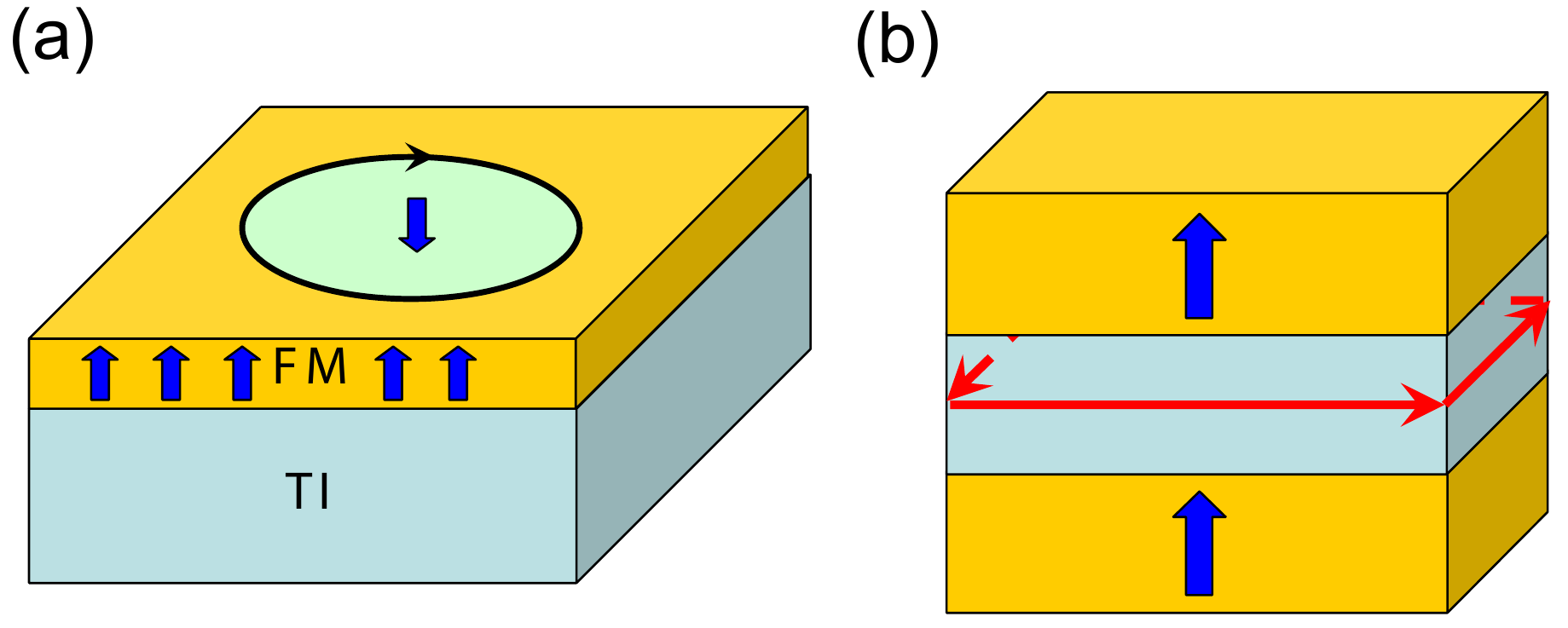}
\end{center}
\caption{(color online) (a) FM layer on the surface of the TI with a magnetic domain wall, along which a chiral edge state propagates.
(b) Illustrate of the QAH effect in ferromagnets-TI heterostructure with parallel magnetization. The magnetic domain wall in (a) is topologically equivalent to (b). The net Hall conductance is given by the summation of the contributions from the top and bottom surfaces. With opposite surface normal vector between top and bottom surfaces, the top and bottom surfaces both contribute $e^2/2h$ Hall conductance. The chiral edge state is trapped on the side surfaces of the TI and carry the quantized Hall current. Reproduced with permission from \cite{qi2011,qi2008}.}
\end{figure}

\section{Phase diagram in the thin film of a magnetic TI}
\label{sec3}

As discussed, the crucial criteria to realize the QAH effect is the spin polarized 2D band inversion. Therefore, the thin film of a TI with FM order is one of most promising materials. Here, in this section, we study the phase diagram in this system.

\subsection{Effective model}
The low-energy bands of this system consist of Dirac-type surface states only, for the bulk states are always gapped. The generic form of the effective Hamiltonian is~\cite{qi2008,yu2010,wang2013a}
\begin{eqnarray}\label{model0}
&&\widetilde{\mathcal{H}}_{\mathrm{surf}}(k_x,k_y)+\widetilde{\mathcal{H}}_{\mathrm{Zeeman}}(k_x,k_y)
   \nonumber
   \\
      &&=\left(\begin{array}{cccc}
   0 & iv_F k_- & m(k) & 0\\
   -iv_Fk_+ & 0 & 0 & m(k)\\
   m(k) & 0 & 0 & -iv_Fk_-\\
   0 & m(k) & iv_Fk_+ & 0
   \end{array}\right)
   \nonumber
   \\
   &&+\left(\begin{array}{cccc}
   \Delta & 0 & 0 & 0\\
   0 & -\Delta & 0 & 0\\
   0 & 0 & \Delta & 0\\
   0 & 0 & 0 & -\Delta
   \end{array}\right),
\end{eqnarray}
with the basis of $|t\uparrow\rangle$, $|t\downarrow\rangle$, $|b\uparrow\rangle$ and $|b\downarrow\rangle$, where $t$, $b$ denote the top and bottom surface states and $\uparrow$, $\downarrow$ represent the spin up and down states, respectively. $v_F$ is the Fermi velocity. $\Delta$ is the exchange field along the $z$ axis introduced by the FM ordering. Here, $\Delta\propto \langle S\rangle$ with $\langle S\rangle$ the mean field expectation value of the local spin. The magnetization $M\propto\langle S\rangle_{\mathrm{ave}}$ where $\langle S\rangle_{\mathrm{ave}}$ is the spatial average of $\langle S\rangle$. $m(k)$ describes the tunneling effect between the top and bottom surface states. In the thick slab geometry ($m(k)\approx0$), the top and bottom surface states are well separated spatially. However, with the reduction of the film thickness, $m(k)$ is finite, and to the lowest order in $k$, $m(k)=M-B(k_x^2+k_y^2)$, and $\left|M\right|<\left|\Delta\right|$ guarantees the system is in the QAH state. There is another term which describes the inversion asymmetry between top and bottom surfaces due to the existence of substrate,
\begin{eqnarray}
\widetilde{\mathcal{H}}_{\mathrm{inv}}=\left(\begin{array}{cccc}
   V & 0 & 0 & 0\\
   0 & V & 0 & 0\\
   0 & 0 & -V & 0\\
   0 & 0 & 0 & -V
   \end{array}\right),
\end{eqnarray}
$V$ represents the magnitude of inversion asymmetry.

In terms of the new basis $|+\uparrow\rangle$, $|-\downarrow\rangle$, $|+\downarrow\rangle$, $|-\uparrow\rangle$ with $|\pm\uparrow\rangle=(|t\uparrow\rangle\pm|b\uparrow\rangle)/\sqrt{2}$
and $|\pm\downarrow\rangle=(|t\downarrow\rangle\pm|b\downarrow\rangle)/\sqrt{2}$, the Hamiltonian (10) and (11) of system becomes
\begin{eqnarray}\label{model1}
&&\mathcal{H}_{\mathrm{total}}(k_x,k_y) =
\left(\begin{array}{cc}
\mathcal{H}_+(k) & V\sigma^1\\
V\sigma^1 & \mathcal{H}_-(k)
\end{array}\right),\\
&&\mathcal{H}_{\pm}(k) = v_Fk_y\tau^1\mp v_Fk_x\tau^2+\left(m(k)\pm\Delta\right)\tau^3
\end{eqnarray}
where $\tau^i$, $\sigma^1$ are Pauli matrices.

\subsection{Phase diagram}

\begin{figure}[htbp]
\label{fig2}
\begin{center}
\includegraphics[width=3.3in,angle=0]{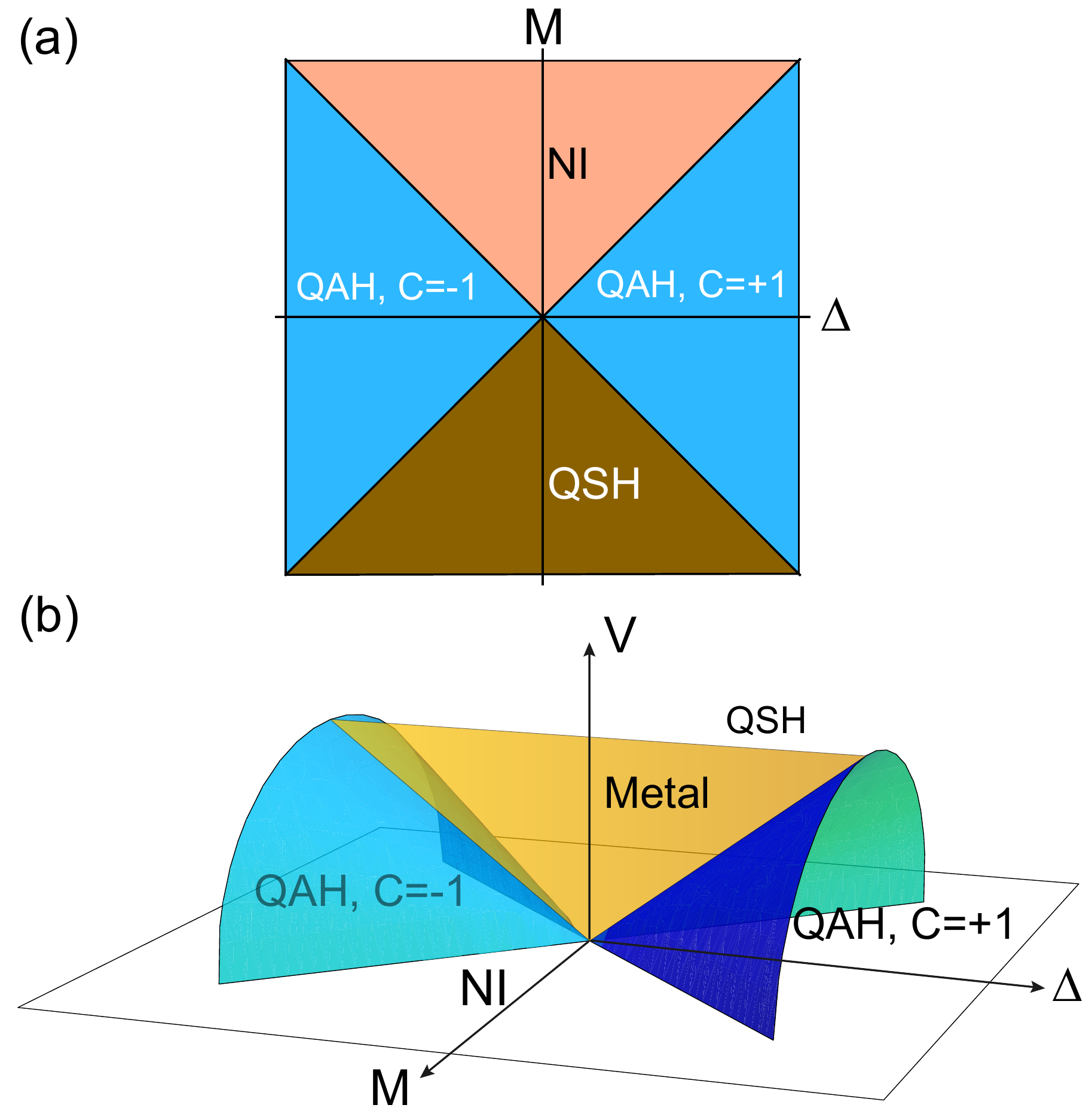}
\end{center}
\caption{(color online) (a) The phase diagram of the thin films of a magnetic TI with $V=0$. The $x$ axis labels the exchange coupling $\Delta$ and $y$ axis labels $M$. Integers $C$ labels the Chern number. (b) Phase diagram for finite $V$ shown only for $V\geq0$. Phase QSH is well defined only in the $\Delta=0$ plane, phase Metal is well defined only in the $M=0$ plane.}
\end{figure}

We will first study the phase diagram of the system in Eq.~(\ref{model1}) for $V=0$. The system is decoupled into two models with opposite chirality. $\mathcal{H}_{\pm}$ are similar to the two band model in Eq.~(1). At half filling, $\mathcal{H}_{\pm}(k)$ have Chern number $\mp 1$ or $0$ depending on whether the Dirac mass is inverted ($m(k)\pm\Delta<0$) or not ($m(k)\pm\Delta>0$) at $\Gamma$ point. Thus the total Chern number of the system is $C=\Delta/|\Delta|$ when $\left|\Delta\right|>\left|M\right|$, and $C=0$ when $\left|\Delta\right|<\left|M\right|$. The Chern number changes by $1$ at $\Delta=\pm M$.

The bulk spectrum is $E_{\mathbf{k}}=\pm\sqrt{v_F^2k^2\pm[m(k)\pm\Delta]^2}$.
Since topological invariants cannot change without closing the bulk gap, the phase diagram can be determined by first finding the phase boundaries which are gapless regions in the $(\Delta,M)$ plane, and then calculating the Chern number of the gapped phases. Assuming $B<0$, then for this model the
critical lines are determined by $|M\pm\Delta|=0$ which leads to the
phase diagram as shown in Fig.~2(a). As expected, the phase boundary reduces to
the critical point $\Delta=0$ between the QSH phase and a trivial or normal insulator (NI) phase in the limit $\Delta=0$. The point
$\Delta=0$ is a multicritical point in this phase diagram. For $M>0$ and $|M|>|\Delta|$ the system is adiabatically connected to a
trivial insulator state with a full gap. For $M<0$ and $\Delta=0$ the system is in a nontrivial QSH state.

Next, we consider the $V\neq0$ in the Hamiltonian in Eq.~(\ref{model1}). Similar to the $V=0$ case, we determine the phase boundaries by the gapless regions in the energy spectrum, which leads to the following condition
\begin{equation}
M^2+V^2 = \Delta^2.
\end{equation}
The entire phase diagram in the $(M,V,\Delta)$ space is shown in Fig.~2(b). Except for the metallic phase in the $M=0$ plane with $|V|>|\Delta|$ and the phase boundaries, there are three gapped phases. The Chern number of each phase can be determined by its adiabatic connection to the $V=0$ limit.

\section{Ferromagnetic order in a magnetically doped TI}
\label{sec4}

As discussed in Section~\ref{sec2}, the key point to realize the QAH effect is to introduce a long-range FM order into the TI. In most conventional diluted magnetic semiconductors, the  Ruderman-Kittel-Kasuya-Yosida (RKKY) mechanism is believed to have induced the long-range FM order, which is the FM coupling between far away magnetic
impurities mediated by itinerant electrons. However, to exhibit the QAH effect, the
FM needs to be present also in the insulating regime. There are two possible mechanism, surface electron mediated
RKKY interaction~\cite{liu2009} and bulk electron mediated Van Vleck paramagnetism~\cite{yu2010}.

\subsection{Surface electron mediated RKKY interaction}

The low-energy effective Hamiltonian of the single Dirac cone surface band is described in Eq.~(\ref{surfaceH}).
The exchange coupling between impurity spin and surface states takes the form
\begin{equation}\label{exchange}
\hat{H}_{\mathrm{int}}=\sum_{i}J_i\mathbf{S}_i\cdot\psi^{\dag}\boldsymbol{\sigma}\psi(\mathbf{R}_i),
\end{equation}
where $J_i$ is the exchange coupling strength, $\mathbf{S}_i$ is the spin of a magnetic impurity, and $\psi^{\dag}\boldsymbol{\sigma}\psi(\mathbf{R}_i)$ is the surface band electron spin density at the magnetic impurity position $\mathbf{R}_i$. The surface electron mediated RKKY interaction between impurities can be obtained by integrating out the fermions in the Hamiltonians (\ref{surfaceH}) and (\ref{exchange}), which results in the form for any two magnetic impurities $S_1$ and $S_2$,
\begin{equation}
\hat{H}_{\mathrm{in}} = \sum\limits_{i,j=x,y,z}\Phi_{i,j}(\left|\mathbf{r}-\mathbf{r}'\right|)\mathbf{S}_{1i}(\mathbf{r})\mathbf{S}_{2j}(\mathbf{r}').
\end{equation}
$\Phi_{i,j}(R)$ is a function of $R=\left|\mathbf{r}-\mathbf{r}'\right|$. When the surface state has a finite Fermi wave vector $k_F$ as is the case in a conventional Fermi liquid, the sign of the RKKY interaction between the impurity spins oscillates with Fermi wavelength $\lambda_F=1/k_F$. If the Fermi level is around the Dirac point so that $k_F\rightarrow0$, the RKKY interaction does not change sign, as shown in Fig.~3. The sign of the resulting nearly uniform spin-spin interaction is determined by a second order perturbation calculation, which is generally FM. It is easy to understand such a FM interaction instead of an antiferromagnetic interaction physically, since a uniform spin polarization opens up a maximal
gap on the surface and minimizes the total kinetic energy. Such a FM spin-spin interaction at the Fermi level of the Dirac point enables
the system to develop a ferromagnetic order spontaneously~\cite{liu2009}.

\begin{figure}[t]
\label{fig3}
\begin{center}
\includegraphics[width=3.0in,angle=0]{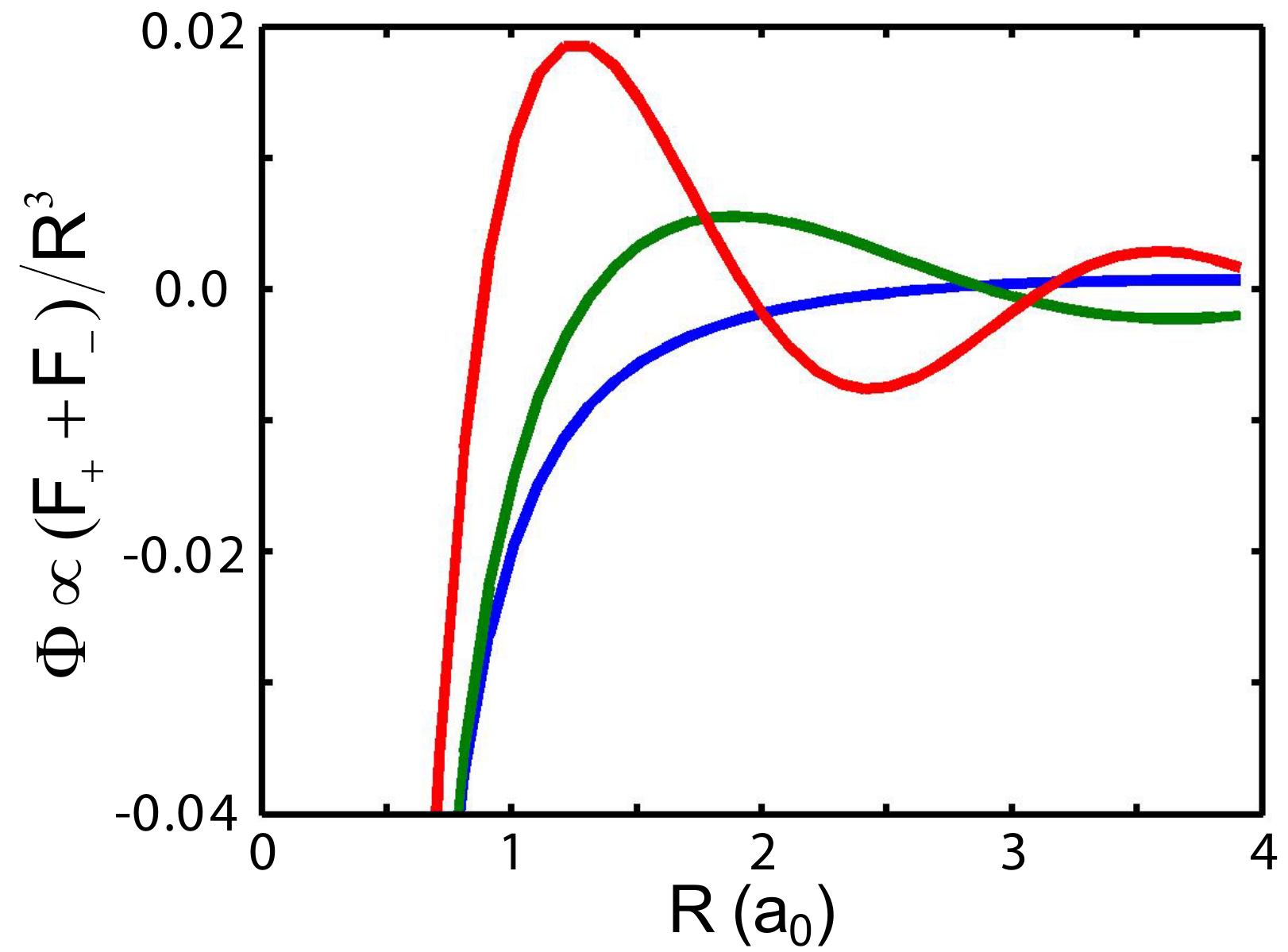}
\end{center}
\caption{RKKY interaction versus the distance $R$ between two magnetic impurities. Fermi momentum $k_F$ for blue, green, and red line is $0.5/a_0$, $1.0/a_0$, and $1.5/a_0$. $a_0$ is the lattice constant. Reproduced with permission from~\cite{liu2009}.}
\end{figure}

\subsection{Ferromagnetism mediated by bulk electrons}
Besides the surface electron mediated RKKY interaction, the magnetic exchange coupling between local moments can also be mediated by insulating bulk electrons through van Vleck paramagnetism. In order to have a nonzero FM transition temperature, a sizable electron spin susceptibility $\chi_e$ is needed~\cite{yu2010}. However, in most dilute magnetic semiconductors, for example, (Ga$_{1-x}$Mn$_x$)As, the electron spin susceptibility is negligible for the insulator phase.

For temperatures much less than the band gap, the spin susceptibility for a band
insulator can be obtained by the second order perturbation on the ground state, which
can be written as
\begin{equation}
\chi^{zz}_e = \sum\limits_{E_{nk}<E_F,E_{mk}>E_F}4\mu_0\mu_B^2\frac{\langle nk|\hat{S}_z|mk\rangle\langle mk|\hat{S}_z|nk\rangle}{E_{mk}-E_{nk}},
\end{equation}
where $\mu_0$ is the vacuum permeability, $\mu_B$ is the Bohr magneton, $E_F$ is the Fermi energy, $\hat{S}_z$
is the spin operator of electron, $|mk\rangle$ and $|nk\rangle$ are the Bloch functions in conduction and
valence bands respectively. In order to have a sizable $\chi_e$, nonzero matrix element of spin operator, $\hat{S}_z$,
between the valance and conduction bands is need. Fortunately, in magnetically doped Bi$_2$Se$_3$ family TIs, the conduction band and valence band have similar electronic orbitals because of the
band inversion, i.e., the bonding and antibonding $p$ orbitals, which results in a considerable van Vleck magnetic susceptibility even when the Fermi energy is in gap. Such a mechanism is absent in GaAs system, which has $s$-like conduction and $p$-like valence bands. Hence the spins
of the magnetic impurities in Bi$_2$Se$_3$ family TIs can be ferromagnetically
coupled via the van Vleck mechamism, even the system is insulating.

\section{Experimental observation of the QAH effect}
\label{sec5}

In a beautiful experiment, the QAH effect has been discovered in Cr$_{0.15}$(Bi$_{0.1}$Sb$_{0.9}$)$_{1.85}$Te$_{3}$ magnetic TI with a thickness of five quintuple layers (QLs) by Xue's group~\cite{chang2013b}. With this composition, the
film is nearly charge neutral~\cite{zhang2011} so that the chemical potential can be fine-tuned to the magnetically induced gap by the back gate.

\begin{figure}[b]
\label{fig4}
\begin{center}
\includegraphics[width=3.4in,angle=0]{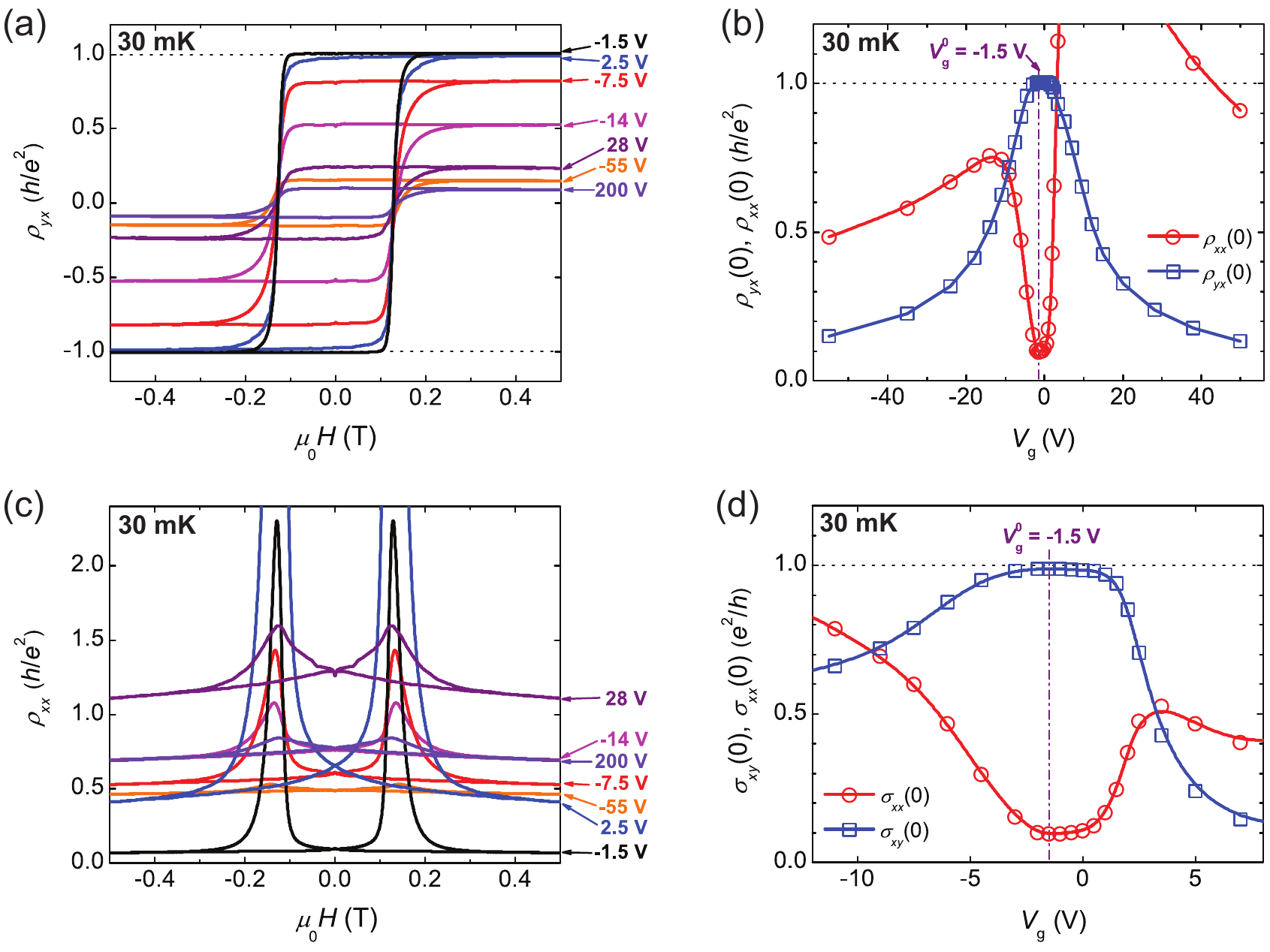}
\end{center}
\caption{The QAH effect observed at $30$~mK. (a) $\rho_{yx}$ as a function of magnetic field at different $V_gs$.
(b) Dependence of $\rho_{yx}(0)$ (empty blue squares) and $\rho_{xx}(0)$ (empty red circles) on $V_g$. (c)  $\rho_{xx}$ as a function of magnetic field at different $V_gs$. (d) Dependence of $\sigma_{xy}(0)$ (empty blue squares) and $\sigma_{xx}(0)$ (empty
red circles) on $V_g$. The vertical purple dashed-dotted lines in (b) and (d) indicate the in-gap gate voltage. Reproduced with permission from~\cite{chang2013b}.}
\end{figure}

Fig.~4, (a) and (c), shows the magnetic field
dependence of $\rho_{yx}$ and $\rho_{xx}$, respectively, measured
at $T=30$~mK at different bottom-gate
voltages ($V_gs$). The shape and coercivity of the
$\rho_{yx}$ hysteresis loops vary little with $V_g$,
thanks to the robust ferromagnetism probably
mediated by the van Vleck mechanism.
In the magnetized states, $\rho_{yx}$ is nearly independent
of the magnetic field, suggesting perfect
FM ordering and charge neutrality of
the sample. On the other hand, the anomalous Hall resistance
(height of the loops) changes dramatically
with $V_g$, with a maximum value of $h/e^2$ around
$V_g=-1.5$~V. The magnetoresistance (MR) curves
[Fig. 4(c)] exhibit the typical shape for a ferromagnetic
material: two sharp symmetric peaks at the coercive fields.
The $V_g$ dependences of $\rho_{yx}$ and $\rho_{xx}$ at zero
field [labeled $\rho_{yx}(0)$ and $\rho_{xx}(0)$, respectively] are
plotted in Fig.~4(b). The most important observation is that the Hall resistance
exhibits a clear plateau with exactly the value
$h/e^2$ at zero magnetic field, where the center of the plateau is at the gate voltage
$V_g=-1.5$~V. This indicates the QAH effect has been realized experimentally. Accompanying the quantization in $\rho_{yx}(0)$,
the longitudinal resistance $\rho_{xx}(0)$ exhibits a sharp
dip down to $0.098 h/e^2$.

\begin{figure}[t]
\label{fig5}
\begin{center}
\includegraphics[width=3.3in,angle=0]{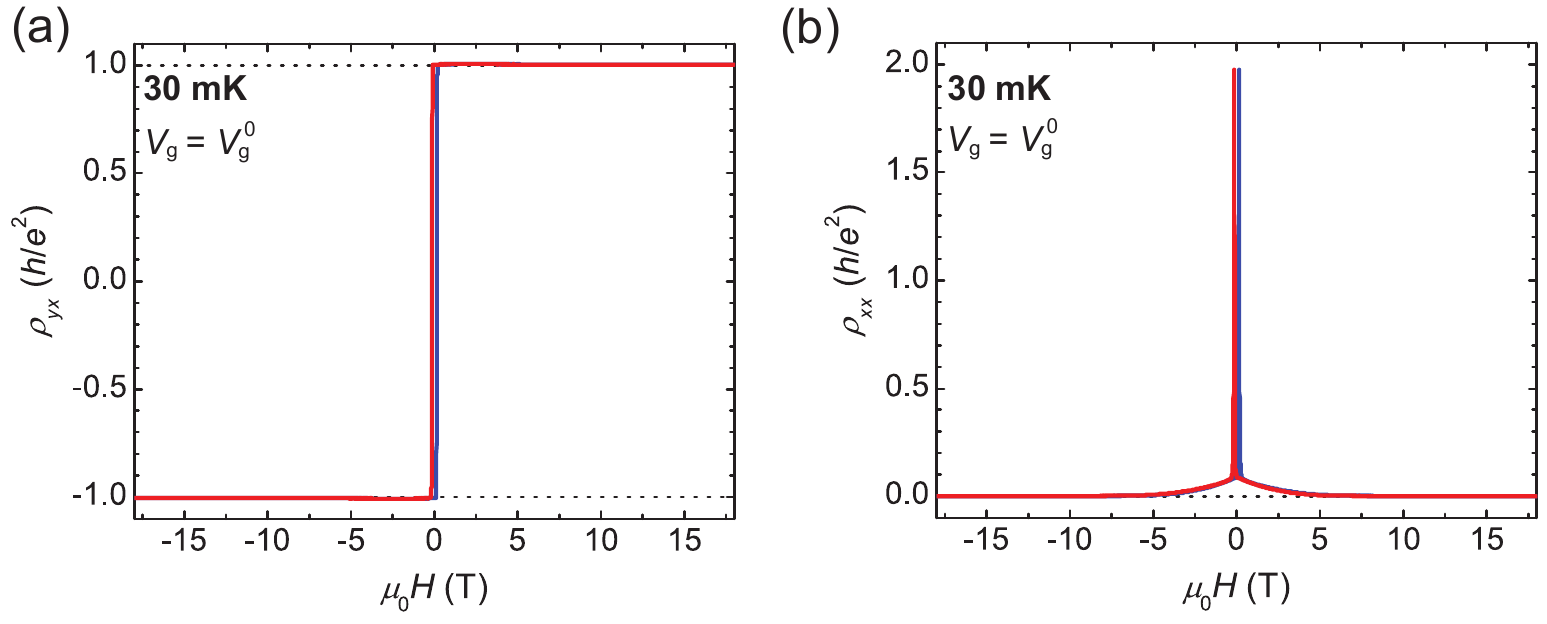}
\end{center}
\caption{The QAH effect under strong magnetic field measured at $30$~mK. (a) Magnetic field
dependence of $\rho_{yx}$ at $V_g^0$. (b) Magnetic field dependence of $\rho_{xx}$ at $V_g^0$. The blue and red lines in (a) and
(b) indicate the data taken with increasing and decreasing fields, respectively. Reproduced with permission from~\cite{chang2013b}.}
\end{figure}

To confirm the QAH effect observed in Fig.~4, they apply
a magnetic field to localize all trivial states that are dissipative
in the sample. The magnetic field dependence of $\rho_{yx}$ and $\rho_{xx}$ of the sample are as shown in Fig.~5(a) and 5(b),respectively.
As the magnetic field increases, the longitudinal resistance $\rho_{xx}$ decreases monotonically after exhibiting a large MR peak at $H_c$. In particular, $\rho_{xx}$ completely vanishes for magnetic field greater than $10$ T, indicating the entrance into a perfect QH regime. Since both $\rho_{xx}$ and $\rho_{xy}$ are smooth functions of the magnetic field (up to $10$ T), there is no quantum phase transition and the sample remains in the same topological phase in the $0-10$ T magnetic field interval, i.e., the QAH phase.

\section{Recent theoretical development of QAH effect}
\label{sec6}

\subsection{QAH effect with higher plateaus}
The topological phases of 2D insulators without symmetry protection are classified by the first Chern number $C\in\mathbb{Z}$, which indicates the presence of $C$ dissipationless chiral edge states. Accordingly, the Hall resistance is quantized into $h/Ce^2$ plateaus. Such plateaus has long been observed in the integer QHE under strong magnetic fields. In the absence of magnetic fields, QAH insulators can also be classified into topological phases with various Chern numbers $C$, while so far only the $C=1$ QAH effect has been observed in experiments~\cite{chang2013b,checkelsky2014,kou2014}. QAH insulators with a higher Chern number, however, could be significant both practically and fundamentally. While the edge channel of QAH insulator is proposed as interconnects for integrated circuits, a QAH insulator with multiple edge channels can greatly reduce the contact resistance. Novel topological phases may also arise when interaction or fractional filling is introduced into the system~\cite{wang2013a}.

\begin{figure}[b]
\label{fig6}
\begin{center}
\includegraphics[width=3.2in,angle=0]{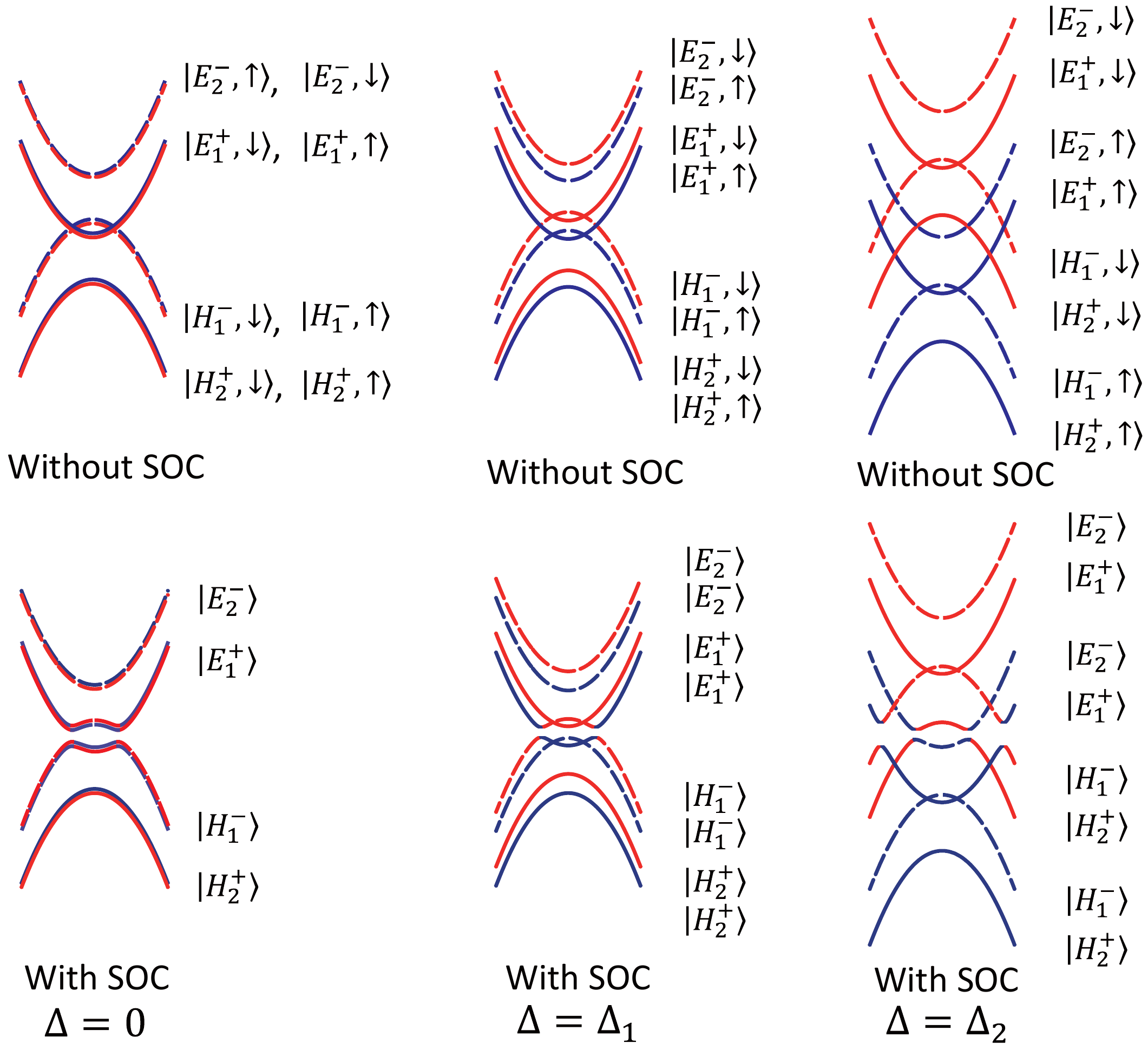}
\end{center}
\caption{Evolution of the subband structure upon increasing the exchange field. The solid lines denote the sub-bands that have even parity at $\Gamma$ point, and dashed lines denote sub-bands with odd parity at $\Gamma$ point. The blue color denotes the spin up electrons; red, spin down electrons. (a) The initial (E$_1$, H$_1$) sub-bands are already inverted, while the (E$_2$, H$_2$) subbands are not inverted. The exchange field $\Delta_1$ release the band inversion in one pair of (E$_1$, H$_1$) subbands and increase the band inversion in the other pair, while the (E$_2$, H$_2$) subbands are still not inverted. With stronger exchange field $\Delta_2$, a pair of inverted (E$_2$, H$_2$) subbands appears, while keeping only one pair of (E$_1$, H$_1$) subbands inverted. Reproduced with permission from~\cite{wang2013a}.}
\end{figure}

With strong enough FM ordering and SOC, the QAH effect with higher Chern numbers can be realized in thin films of magnetic TIs. The magnetic TIs can be made out of Bi$_2$Te$_3$, Bi$_2$Se$_3$ or Sb$_2$Te$_3$ compounds dopped with Cr or V. The basic mechanism is to have \emph{multiple pairs of inverted spin-polarized bands}, each of which contributes a Chern number $1$ or $-1$. Their sum gives the total Chern number $C$ of the system, which can in principle be any integer.

Consider a 2D thin film of dopped TIs with a spontaneous FM order. The low-energy bands consist of a bonding and an antibonding state of $p_z$ orbitals, denoted by
$\left|P2^-_z,\uparrow(\downarrow)\right\rangle$ and $\left|P1^+_z,\uparrow(\downarrow)\right\rangle$, respectively. The 3D effective Hamiltonian describing these four bands is
\begin{equation}\label{model}
\mathcal{H}_{\mathrm{3D}}(k_x,k_y,k_z)
   =\left(\begin{array}{cc}
   H_+(k) & A_1k_zi\sigma_y\\
   -A_1k_zi\sigma_y & H_-(k)
   \end{array}\right),
\end{equation}
\begin{equation}
H_{\pm}(k)=\varepsilon(k)+d^i_{\pm}(k)\tau_i\nonumber\ .
\end{equation}
where $\tau_i$ ($i=1,2,3$) and $\sigma_y$ are Pauli matrices, $d^{1,2,3}_{\pm}(k)=(A_2k_x,\pm A_2k_y,M(k)\mp\Delta)$. To the lowest order in $k$, $M(k)=B_0+B_1k_z^2+B_2(k_x^2+k_y^2)$, $\varepsilon(k)=D_0+D_1k_z^2+D_2(k_x^2+k_y^2)$ accounts for the particle-hole asymmetry. $B_0<0$ and $B_1, B_2>0$ guarantee the system is in the inverted regime. The basis of Eq.~(\ref{model}) is $(|P1^{+}_z,\uparrow\rangle$, $|P2^{-}_z,\downarrow\rangle$, $|P1^{+}_z,\downarrow\rangle$, $|P2^{-}_z,\uparrow\rangle)$, where the $\pm$ in the basis stand for the even and odd parity and $\uparrow$, $\downarrow$ represent spin up and down states, respectively. $\Delta$ is the exchange field along the $z$ axis introduced by the FM ordering.

For thin films with a thickness $d$, $k_z$ becomes quantized, and the $z$-direction wave function takes the form $\varphi_n(z)=\sqrt{2/d}\sin (n\pi z/d)$ ($n\in\mathbb{Z}^+,z\in[0,d]$), leading to a series of 2D subbands labeled by index $n$. In the limit $A_1=0$, the Hamiltonian is decoupled into many two-band 2D models $h_+(n)$ and $h_-(n)$ with opposite chirality:
\begin{equation}\label{Hn}
\tilde{\mathcal{H}}_{\mathrm{2D}}(n)
  =\left(\begin{array}{cc}
  h_+(n) & 0\\
  0 & h_-(n)
  \end{array}\right)\ ,
\end{equation}
where $h_{\pm}(n)=\tilde{\varepsilon}_n1_{2\times2}+(\tilde{M}_n\mp\Delta)\tau_3+A_2k_x\tau_1\pm A_2k_y\tau_2$, with the notations $\tilde{\varepsilon}_n=D_0+D_1(n\pi/d)^2+D_2(k_x^2+k_y^2)$, $\tilde{M}_n=B_0+B_1(n\pi/d)^2+B_2(k_x^2+k_y^2)$. The basis for $\tilde{\mathcal{H}}_{\mathrm{2D}}(n)$ is given by $(|E_n,\uparrow\rangle,|H_n,\downarrow\rangle,|E_n,\downarrow\rangle,|H_n,\uparrow\rangle)=\varphi_n(z)(|P1^{+}_z,\uparrow\rangle$, $|P2^{-}_z,\downarrow\rangle$, $|P1^{+}_z,\downarrow\rangle$, $|P2^{-}_z,\uparrow\rangle)$. When half of the bands are filled, each model $h_\pm(n)$ has a Chern number $\pm1$ or $0$, depending on whether the Dirac mass is inverted ($\tilde{M}_n\mp\Delta<0$) or not ($\tilde{M}_n\mp\Delta>0$) at $\Gamma$ point. The total Chern number of the system becomes
\begin{equation}\label{chern}
C = N_+-N_-\ ,
\end{equation}
where $N_{\pm}$ is the number of $h_{\pm}(n)$ with inverted Dirac mass, respectively. In the non-magnetic case when $\Delta=0$, $N_+=N_-$, and one gets either the trivial insulator or the quantum spin Hall (QSH) insulator. When $\Delta\neq0$, $N_+$ can be different from $N_-$, and QAH insulators with various Chern numbers can be realized (Fig.~6). In the case $A_1=0$ discussed here, neglecting the particle-hole-asymmetric term $\tilde{\varepsilon}_n$, the band inversion of $h_{\pm}(n)$ takes place when $d=n\pi\sqrt{B_1/(\pm\Delta-B_0)}$. These equations give the phase boundaries between topological phases with different Chern numbers (Fig.~7(a)). As is seen, the Chern number of the system tends to be larger as the exchange field $\Delta$ and the thickness $d$ increases (Fig.~7(a)). When $\Delta\ge |B_0|$, the Chern number monotonically increases as a function of the thickness $d$.

\begin{figure}[b]
\label{fig7}
\begin{center}
\includegraphics[width=3.0in,angle=0]{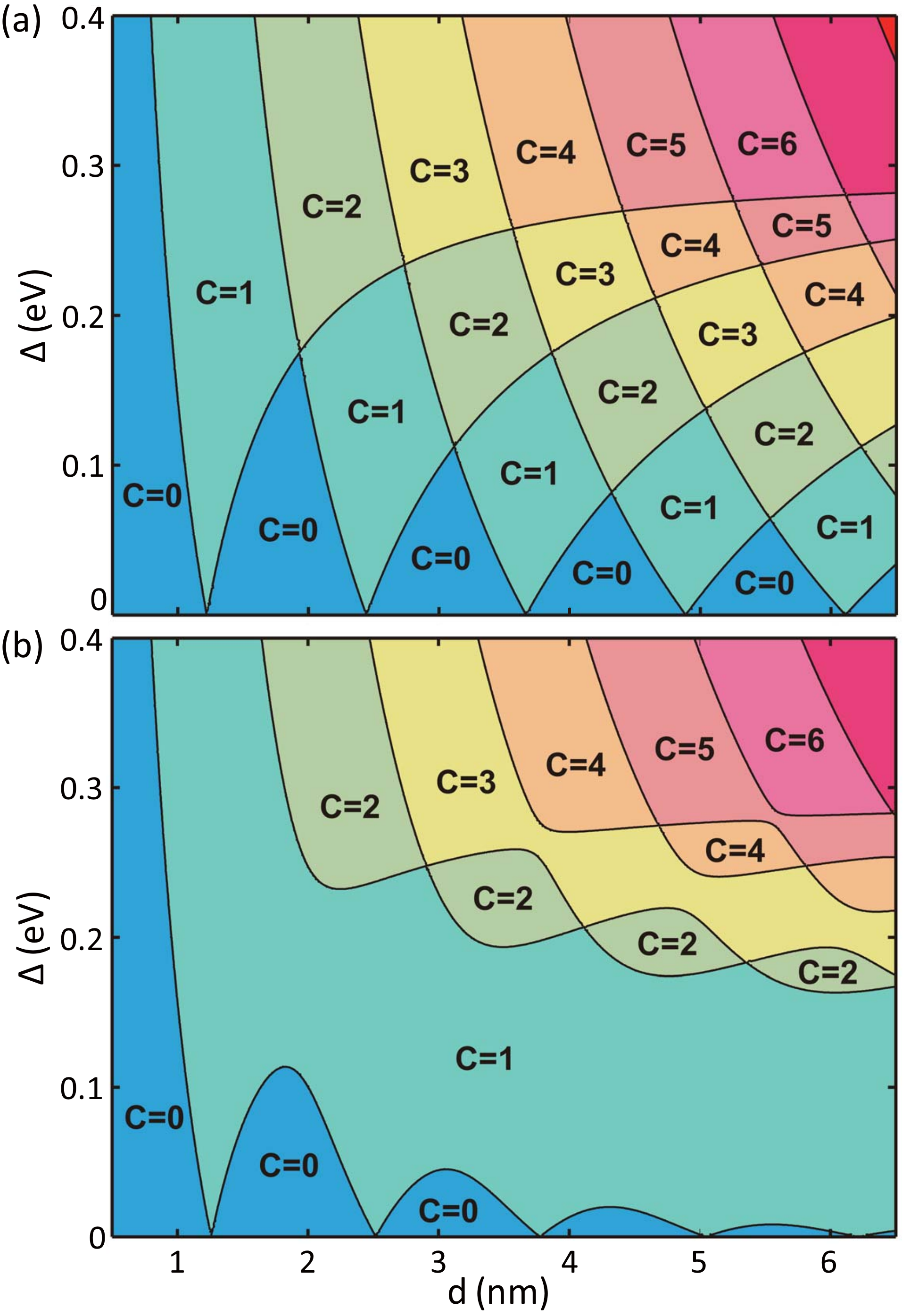}
\end{center}
\caption{The phase diagram of QAH effect in thin films of magnetic TIs with
two variables: the exchange field $\Delta$ and thickness of thin film $d$. All parameters are taken from Ref.~\cite{zhang2009} for Bi$_2$(Se$_{0.4}$Te$_{0.6}$)$_3$. (a) and (b) are phase diagrams without and with $A_1$ term, respectively. The different QAH phases are denoted by corresponding Chern numbers. The particle-hole asymmetric term is neglected for it does not change the topology of the phase diagram. The width of each Hall plateau in QAH effect depends on the band parameters and the thickness of the material, which is distinct from that in integer QHE. Reproduced with permission from~\cite{wang2013a}.}
\end{figure}

When the $A_1$ term is turned on, it induces a coupling between $h_{\pm}(n)$ and $h_{\mp}(n+1)$. As is shown in (Fig.~7(b)), this makes the phase spaces of odd Chern number phases enlarged and united, while those of even Chern numbers separated into ``islands''. In particular, a large phase space is occupied by the $C=1$ phase. The realization of QAH effect with higher plateaus thus requires a large enough exchange field.

To realize QAH effect with a Chern number $C>1$, it is necessary to have a large enough exchange field $\Delta$ or sufficiently many low energy 2D subbands (large enough thickness). The exchange field $\Delta$ can be enhanced by increasing the magnetic elements doping concentration; however, the doping concentration cannot be too high, or the structure of the material will be totally changed. The optimal way is to tune the doping concentration of the material to the vicinity of topological phase transition point, where the 3D bulk band gap is small, thus the FM exchange coupling would make more pairs of inverted 2D spin polarized bands possible. For example, the topological phase transition of Bi$_{2-y}$Cr$_y$(Se$_{0.6}$Te$_{0.4}$)$_3$ is shown to happen at $y=0.22$~\cite{zhang2013}. Correspondingly, the thin film of this material will have many low energy 2D subbands~\cite{wang2013a}.

\subsection{Anomalous edge transport in the QAH effect}
\label{transport}

In the integer QHE with $C$ Landau levels filled, the Hall resistance is quantized into $\rho_{xy}=h/Ce^2$, while the longitudinal resistance $\rho_{xx}$ vanishes exactly. Theoretically, the QAH effect with solely $C$ chiral edge states should show similar transport properties at zero magnetic field. However, in all experimental observation of the QAH effect ($C=1$) in thin films of Cr doped magnetic TIs~\cite{chang2013b,checkelsky2014,kou2014}, the Hall resistance ($\rho_{xy}$) reaches a quantized value $h/e^2$ while a small non-zero residual longitudinal resistance ($\rho_{xx}$) remains [see Fig.~4(b)]. Though the Hall resistance is consistent with the quantum transport of a single chiral edge state, the non-vanishing longitudinal resistance indicates the existence of dissipation in the system. This dissipation can be explained by additional (trivial) quasi-helical edge states that coexists with the chiral edge state~\cite{wang2013b}. These nonchiral edge states are not immune to backscattering, and therefore contributes to the dissipative transport. The resistance of such a system exhibits non-Ohmic behavior, and gives nearly quantized $\rho_{xy}$ while having non-zero $\rho_{xx}$.

The coexistence of nonchiral edge states with the chiral edge state can generally happen in the experiment of QAH effect. For five QLs of Cr$_x$(Bi,Sb)$_{2-x}$Te$_3$ studied in the experiment~\cite{chang2013b}, it is shown by first principles calculations that there are nonchiral edge states coexisting with the chiral edge state~\cite{wang2013b}. The nonchiral edge states originates from helical edge states of the QSH effect with TR symmetry broken by the magnetic moments, where the gap of the helical edge states is opened at the Dirac point and buried in the valence band. They can therefore be dubbed as the quasi-helical edge states. Due to TR symmetry breaking, they are no longer immune to the backscattering, and become dissipative.

\begin{figure}[htbp]
\label{fig8}
\begin{center}
\includegraphics[width=3.3in,angle=0]{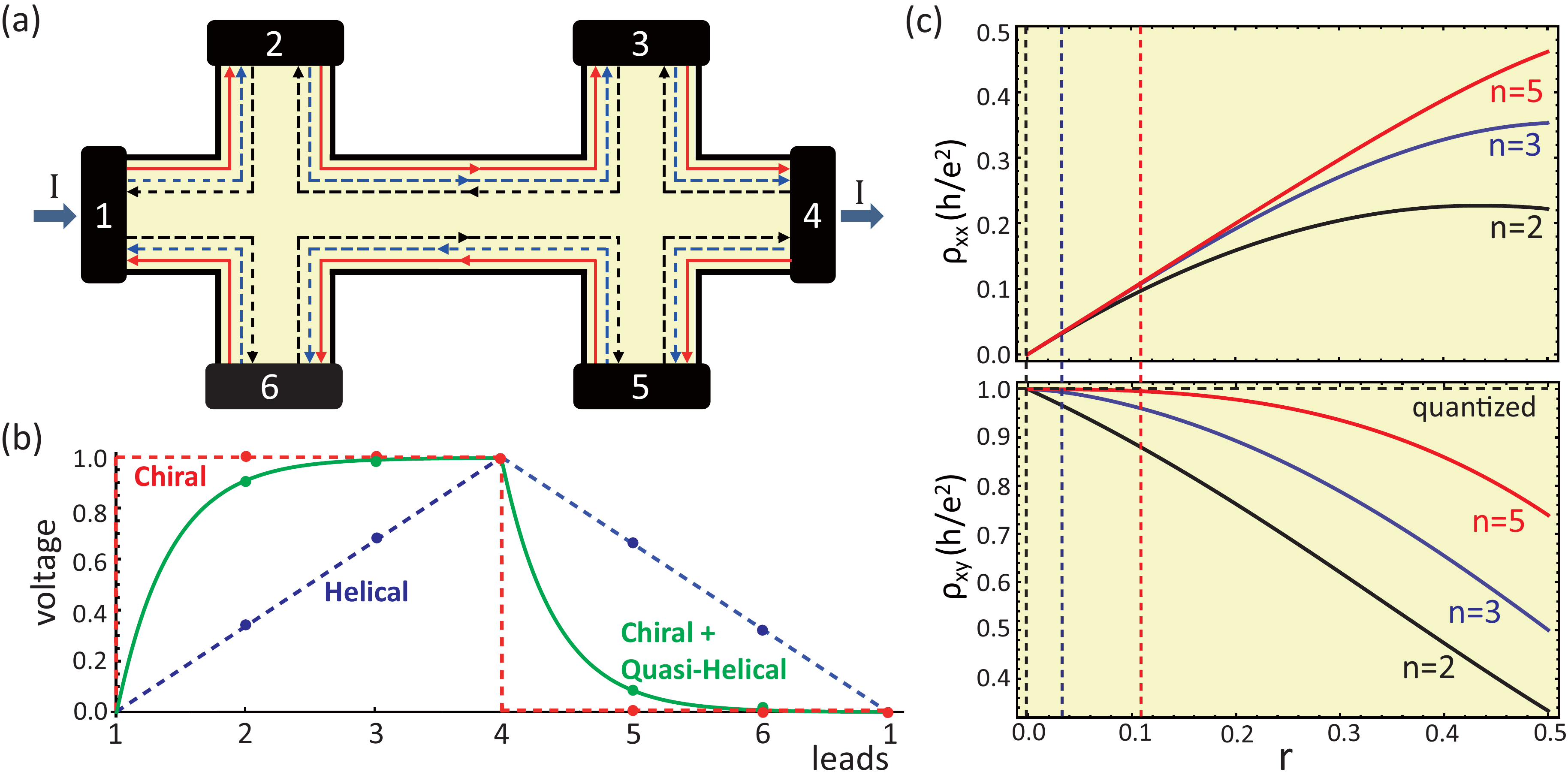}
\end{center}
\caption{(color online) Hall bridge and transport properties. (a) Schematic drawing of a Hall bar device of QAH effect with quasi-helical edge channels (blue and black dashed) coexisting with chiral edge channel (red solid). The current is from terminal 1 to 4. (b) Voltage at terminal 1-6. The QAH effect with coexistence of chiral and quasi-helical edge channels (green) show non-ohmic behaviors of $\rho_{xx}$. (c) $\rho_{xx}$ and $\rho_{xy}$ vs. $r$ with different numbers of effective voltage leads on each side of the sample. Reproduced with permission from~\cite{wang2013b}.}
\end{figure}

The edge transport of such a system can be studied by the Landaur-B\"{u}ttiker formalism. The general relationship between the currents and voltages is expressed as
\begin{equation}\label{LBformula}
I_i=\frac{e^2}{h}\sum\limits_{j}\left(T_{ji}V_i-T_{ij}V_j\right),
\end{equation}
where $V_i$ is the voltage on the $i$th electrode, $I_i$ is the current flowing from the $i$th electrode into the sample, and $T_{ji}$ is the transmission probability from the $i$th to the $j$th electrode. There is no net current ($I_j=0$) on a voltage lead or floating probe $j$, and the total current is conserved, $\sum_iI_i=0$. The currents are zero when all the potentials are equal, implying the constraint $\sum_iT_{ji}=\sum_iT_{ij}$.

For a standard Hall bar with $\mathcal{N}$ current and voltage leads [such as Fig.~8(a) with $\mathcal{N}=6$], the transmission probability for the chiral state of the $\nu=1$ QAH effect are given by $T_{i+1,i}=1$, for $i=1,...,\mathcal{N}$ ($\mathcal{N}+1$ identified with $1$), and others $=0$. For quasi-helical states, $T_{i+1,i}=k_1$, $T_{i,i+1}=k_2$ and others $=0$, where $k_1, k_2<1$ since they are dissipative. In general, $k_1$ and $k_2$ become zero for infinitely large sample. Thus the nonzero total transmission matrix elements are
\begin{equation}\label{transmission}
T_{i+1,i} = 1+k_1,\ \ T_{i,i+1} = k_2.
\end{equation}
In the case of current leads on electrodes 1 and 4, and voltage leads on electrodes 2, 3, 5, and 6 (see Fig.~8(a)), one finds that $I_1=-I_4\equiv I$. Setting $V_1\equiv 0$ and $V_4\equiv V$, and $r\equiv k_2/\left(1+k_1\right)$, one finds the following solution:
\begin{eqnarray}
V_j &=& \frac{1-r^{j-1}}{1-r^3}V, \ \ 1 \leq j\leq 4,
\\
V_j &=& \frac{1-r^{j-7}}{1-r^{-3}}V, \ \ 4 \leq j\leq 6,
\end{eqnarray}
while $I$ is given by Eq. (\ref{LBformula}). In the case of pure chiral edge state transport in QAH effect where $k_1=k_2=0$, one finds $\rho_{xy}\equiv\left(V_2-V_6\right)/I=h/e^2$ and $\rho_{xx}\equiv\left(V_3-V_2\right)/I=0$ as expected. For the helical edge state transport in QSH effect with $T_{i+1,i}=T_{i,i+1}=1$, $R_{14,14}\equiv\left(V_4-V_1\right)/I=3h/2e^2$ and $R_{14,23}\equiv\left(V_3-V_2\right)/I=h/2e^2$, and one has the Ohm's law. When there are both chiral and quasi-helical states, the voltages of different leads vary exponentially, as is shown in Fig.~8(b), and $\rho_{xx}$ does not scale linearly with the spacing between the voltage leads as Ohm's law. Moreover, $\rho_{xx}$ is nonzero while $\rho_{xy}$ is nearly quantized. In the experiment where the sample is usually large, the effect of decoherence between two real leads can be modeled by an extra floating lead, which destroys the coherence by introducing infinitely many low-energy degrees of freedom. Therefore, the standard Hall bar in Fig.~8(a) has effectively $n=5$ voltage leads on each side. For certain parameter range of $r$, $\rho_{xy}$ can be quantized to $h/e^2$ plateau whereas $\rho_{xx}$ is nonzero (see Fig.~8(c)), which explains the dissipative longitudinal transport observed in the QAH effect.

\begin{figure}[t]
\begin{center}
\includegraphics[width=3.3in,angle=0]{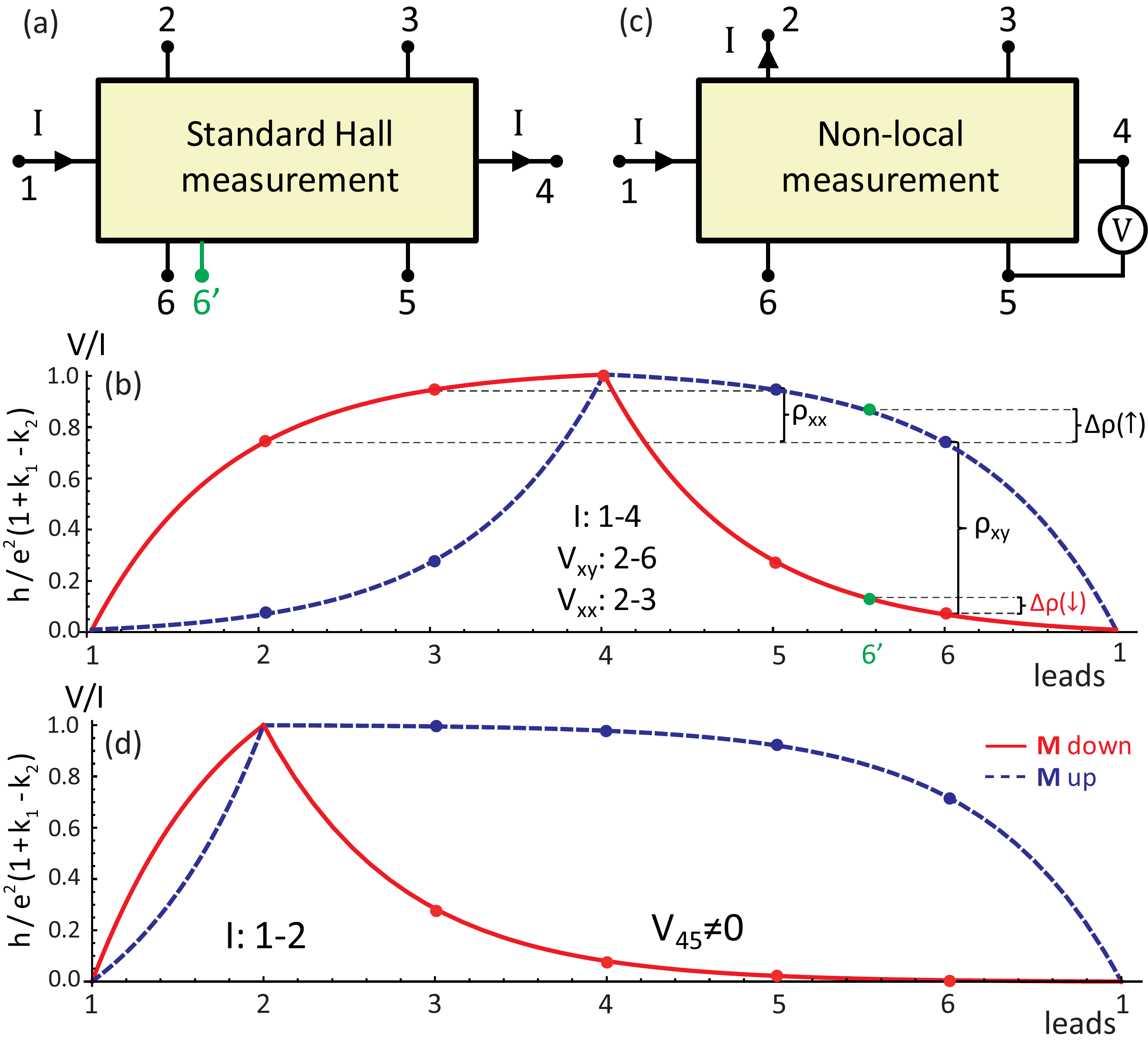}
\end{center}
\caption{(color online) Six-terminal Hall and nonlocal measurements. (a) Standard Hall measurement with six terminals and (b) corresponding voltages.
The current is through 1 to 4, and the Hall voltage is measured between 2 and 6. Terminal $6$ (denoted as $6'$) is not be symmetric to terminal 2 due to misalignment, thus Hall signal may contain some longitudinal component. (c) Nonlocal measurement and (d) voltage. The current is through 1 to 2. The voltage with downward and upward magnetic orderings are denoted as red solid and blue dashed line, respectively. Reproduced with permission from~\cite{wang2013b}.}
\label{fig9}
\end{figure}

In the experiment, the magnetization ${\bf M}$ of the magnetic TI can be flipped to either up ($\uparrow$) or down ($\downarrow$). In the case of pure chiral edge transport, the Hall resistances in these two cases should equal, namely $\rho_{xy}(\uparrow)=-\rho_{xy}(\uparrow)$. It is however observed that $\rho_{xy}(\uparrow)\neq-\rho_{xy}(\uparrow)$. This can be interpreted with the above picture as follows. In reality, the voltage leads may not be aligned very symmetrically. As is shown in Fig.~9(a), it is possible that the lower left voltage lead is at position $6'$ instead of $6$, which is mirror reflection of the voltage lead $2$. Fig.~9(b) shows the voltages when the magnetization is up (the blue dashed line) and down (the red solid line). Denote $\rho_{0}=(V_2^\uparrow-V_6^\uparrow)/I=-(V_2^\downarrow-V_6^\downarrow)/I$, and one finds $\rho_{xy}(\uparrow)=\rho^0-\Delta\rho(\uparrow)$ and $\rho_{xy}(\downarrow)=-\rho^0-\Delta\rho(\downarrow)$, where the errors $\Delta\rho(\uparrow)$ and $\Delta\rho(\downarrow)$ are of the same sign. The two Hall resistance are thus not equal to each other. When $\rho_0$ is near the quantized value, one of them is smaller than $h/e^2$ while the other is larger than $h/e^2$, which agrees with the experimental observation.

The existence of nonchiral edge states can be directly detected by non-local transport measurement. As is shown in Fig.~\ref{fig9}(c) and \ref{fig9}(d), the current passes through leads 1 and 2, and voltage is measured between leads 4 and 5 far away from the bulk current path. In the pure chiral edge state case, the voltage is zero. In the case nonchiral edge states coexisting with the chiral edge state, the voltage is nonzero, leading to a nonzero nonlocal resistance $R_{12,45}=(V_4-V_5)/I$. Remarkably, it is easily seen in Fig. \ref{fig9}(d) that the nonlocal resistance depends on the direction of the magnetization, namely
\begin{equation}\label{R1245}
R_{12,45}(\uparrow)\neq R_{12,45}(\downarrow).
\end{equation}
On the contrary, the bulk dissipation has almost no contribution to the nonlocal resistance, and has no dependence on the magnetization direction. The nonlocal measurement can therefore be used to prove the existence of nonchiral edge states. These nonlocal behaviors have been observed in recent experiments of the QAH effect~\cite{kou2014}.

There are also other methods which can be used for proving coexistence of nonchiral edge states and the chiral edge state. For example, by adding extra floating leads to the standard Hall bar, one can increase the decoherence of the nonchiral edge states and reduce the longitudinal resistance $\rho_{xx}$. Another method is to measure the QAH effect in the Corbino geometry, in which case the current cannot flow from the inner ring to the outer ring via edge, and the quantization of Hall resistance will become exact.

\subsection{Universal scaling behavior at the QAH plateau transition}
\label{scaling}

Continuous phase transitions usually exhibits critical scaling behaviors. In the integer QHE plateau transition driven by the strong external magnetic field, $\rho_{xy}$ changes rapidly in a narrow interval of magnetic field, while $\rho_{xx}$ exhibits a peak, reflecting a delocalization transition in the Landau levels. The localization length $\xi$ diverges in a power law $\xi\sim|B-B_c|^{-\nu}$ with a universal critical exponent $\nu$, which is experimentally measured to be $\nu\approx 2.38$~\cite{li2009,koch1991b,engel1993}. This delocalization transition can be approximated by the Chalker-Coddington network model, whose critical exponent is $\nu\approx 2.4\pm0.2$ as is shown by numerical simulations~\cite{chalker1988,lee1993,slevin2009}.

In the QAH effect experiment, the exchange field $\Delta$ can be tuned from positive to negative by an external magnetic field, and a plateau transition from $\rho_{xy}=h/e^2$ to $\rho_{xy}=-h/e^2$ is driven at the coercivity field ~\cite{chang2013b}. Theoretically, it is shown that this QAH plateau transition at the coercivity field is a quantum phase transition and has universal critical behavior~\cite{wang2014a}. In fact, the QAH plateau transition model can be mapped to the Chalker-Coddington network model, though driven by a different mechanism from that of the integer QHE plateau transition. The critical exponent for QAH phase transition is therefore expected to be $\nu\approx 2.4\pm0.2$. In addition, since the Chern number of the system always changes by $1$ at each transition, it is predicted that the transition from $\rho_{xy}=h/e^2$ to $\rho_{xy}=-h/e^2$ should show an intermediate plateau $\rho_{xy}=0$ for low enough temperatures or large enough sample sizes, and the longitudinal resistance should exhibit a double peak during the transition~\cite{wang2014a}.

The $C=1$ QAH effect is most easily governed by the Hamiltonian
\begin{equation}
\mathcal{H}_0(k_x,k_y)=
\left(\begin{array}{cc}
H_+(k) & 0\\
0 & H_-(k)
\end{array}\right),
\end{equation}
\begin{equation}
H_{\pm}(k)=k_y\tau_1\mp k_x\tau_2+\left(m(k)\pm\Delta\right)\tau_3.
\end{equation}
where $\tau_i$ are Pauli matrices, we set $v_F\equiv1$, and $m(k)=M-B(k_x^2+k_y^2)$. The total Chern number of the system is $\Delta/|\Delta|=\pm1$ when $|\Delta|>|M|$, and is $0$ when $|\Delta|<|M|$. When the direction of magnetization of a QAH insulator is tuned from up to down by the external magnetic field, there are two successive plateau transitions at $\Delta=\pm M$. In reality, at the coercivity field where the magnetization flips, the material always consists of many random domains, and the following three kinds of disorder presents:
\begin{eqnarray}
\mathcal{H}_A &=& A_x(x,y)\tau_2\otimes\sigma_3-A_y(x,y)\tau_1\otimes1,\\
\mathcal{H}_{\Delta} &=& \Delta(x,y)\tau_3\otimes\sigma_3,\\
\mathcal{H}_{V} &=& V(x,y),
\end{eqnarray}
where $\sigma_i$ is the pauli matrix, $\mathbf{A}(x,y)\equiv(A_x,A_y)$, $\Delta(x,y)$, and $V(x,y)$ are the random vector potential, exchange field and scalar potential, with their mean values equal to $0$, $\Delta$ and $0$, respectively.

It is sufficient to consider only the transition of $H_{+}(k)$ at $\Delta=-M$, which becomes exactly the random Dirac model if the approximation $B=0$ is made. By a unitary transformation $G=(\tau_2-\tau_3)/\sqrt{2}$, $H_{+}(k)$ can be rewritten in real space as
\begin{equation}
\widetilde{H}_{+}=GH_{+}G^\dag=(-i\partial_x-A_x)\tau_3-(-i\partial_y-A_y)\tau_1-\delta\tau_2+V\ ,
\end{equation}
where $\delta(x,y)=\Delta(x,y)+M$. At low energies, the evolution operator in a unit time is
\begin{equation}\label{evolution}
\mathcal{U}=e^{-i\widetilde{H}_+}\approx 1-i\widetilde{H}_+-\frac{\widetilde{H}_+^2}{2}
\approx e^{-iV}
 \left(\begin{array}{cc}
   \gamma &\alpha \\
   -\alpha^* & \gamma^*
 \end{array}\right),
\end{equation}
where
\begin{eqnarray}
\gamma(x,y) &=& \cos\delta\cos\left(-i\partial_y-A_y\right)e^{-i\left(-i\partial_x-A_x\right)}\ ,\\
\alpha(x,y) &=& e^{i\left(-i\partial_y-A_y\right)}\left[\sin\delta+i\sin\left(-i\partial_y-A_y\right)\right] .
\end{eqnarray}

\begin{figure}[t]
\begin{center}
\includegraphics[width=3.3in,angle=0]{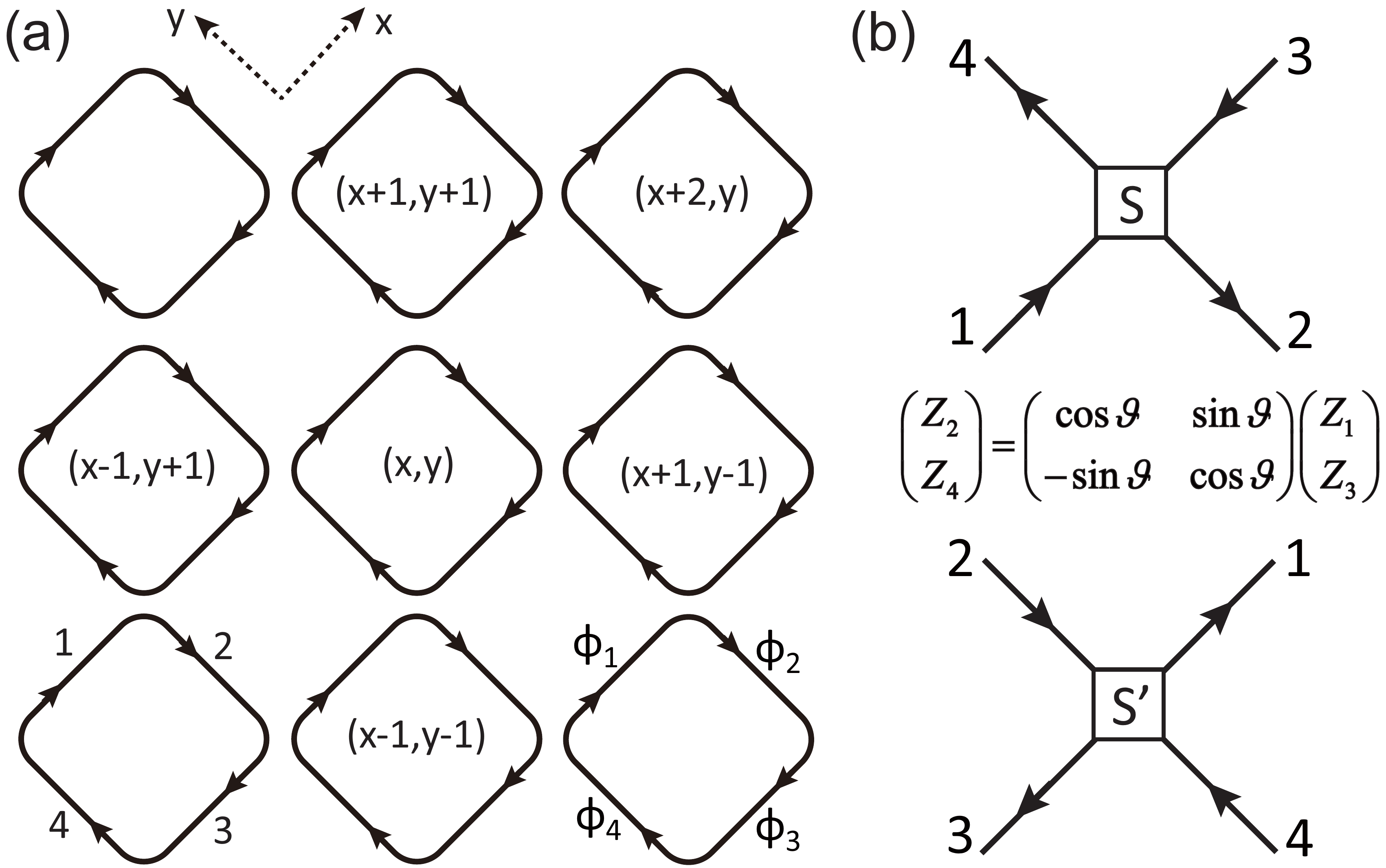}
\end{center}
\caption{The network model. (a) shows the coordinate system for plaquettes and the labeling of the four edges.
(b) Amplitudes associated with possible scattering paths at nodes. Reproduced with permission from~\cite{wang2014a}.}
\label{fig10}
\end{figure}

This matrix can be viewed as a $2$-step scattering matrix of the Chalker-Coddington network model. The network model is defined on a square lattice of plaquettes as is shown in Fig.~10. The four edges of plaquette $(x,y)$ is denoted as $(x,y,i)$ with $i=1,2,3,4$. An electron can propagate along the edges of each plaquette in the direction indicated by the arrow, and can be scattered at the vertices connecting two plaquettes. By denoting the scattering amplitude at edge $(x,y,i)$ as $Z_i(x,y)$, the scattering process can be described with a parameter $\vartheta(x,y)\in[0,\pi/2]$ as
\begin{equation}
\left(\begin{array}{c}Z_2\\Z_4\end{array}\right)=\left(\begin{array}{cc}\cos\vartheta&\sin\vartheta\\-\sin\vartheta&\cos\vartheta\end{array}\right) \left(\begin{array}{c}Z_1\\Z_3\end{array}\right) .
\end{equation}
Besides, the amplitude $Z_i(x,y)$ acquires a phase factor $e^{i\phi_i(x,y)}$ during the propagation along edge $(x,y,i)$. Both $\vartheta(x,y)$ and $\phi_i(x,y)$ are random in space. A scattering matrix $\mathcal{S}(\vartheta,\phi_i,t^x_\pm,t^y_\pm)$ can then be written down in the basis of $\left\{Z_1(x,y), Z_3(x,y); Z_2(x,y), Z_4(x,y)\right\}$, where $t^x_\pm,t^y_\pm$ are the translation operators along $\pm x,\pm y$, respectively. It is easy to show that the two-step scattering matrix takes the form
\begin{equation}
\mathcal{S}^2=\left(\begin{array}{cc}\mathcal{D}_u & \\ & \mathcal{D}_d\end{array}\right) .
\end{equation}
In the continuum limit, the transition operators can be approximated as $t^x_\pm=e^{\pm\partial_x}$ and $t^y_\pm=e^{\pm\partial_y}$, and one finds exactly $\mathcal{D}_u=\mathcal{U}$ by identifying $A_x=(\phi_1-\phi_3)/2$, $A_y=(\phi_4-\phi_2)/2$, $V=-\sum_{i=1}^{4}\phi_i/2$ and $\vartheta=\pi/4+\delta/2$.

The scattering wave function of the network model has a localization length $\xi$ that diverges if $\langle\vartheta\rangle=\vartheta_c=\pi/4$ and $\phi_i$ uniformly distributed in $[0,2\pi]$. Numerical simulations show that $\xi\propto |\langle\vartheta\rangle-\vartheta_c|^{-2.4\pm0.2}$ when deviating from the critical point. When mapped into the plateau transition of the QAH effect, this means the localization length of the electron state at the Fermi level has the scaling behavior $\xi\propto|\langle\delta\rangle|^{-\nu}$, where $\nu\approx2.4$. In the experiment, the exchange field $\Delta$ is proportional to the external magnetic field $H$. Therefore, the localization length can be expressed as $\xi=\xi_0|H-H^*|^{-\nu}$, where $H^*$ is the critical magnetic field for the phase transition.

The longitudinal conductance of an insulating system takes the form $\sigma_{xx}=\sigma_{\max}e^{-L_{\mathrm{eff}}/\xi}$, where $L_{\mathrm{eff}}$ is the effective size. It is therefore expected that
\begin{equation}
\sigma_{xx}(H)=\sigma_{\max}e^{-(L_{\mathrm{eff}}/\xi_0)(H-H^*)^\nu}
\end{equation}
at the plateau transition of the QAH. Similarly, the derivative of $\sigma_{xy}$ with respect to $H$ obeys a similar law,
\begin{equation}
\frac{\partial\sigma_{xy}}{\partial H}(H)=\frac{\nu}{2\Gamma(1/\nu)}\frac{e^2}{h}\left(\frac{L_{\mathrm{eff}}}{\xi_0}\right)^{1/\nu}e^{-(L_{\mathrm{eff}}/\xi_0)(H-H^*)^\nu}\ ,
\end{equation}
where $\Gamma(1/\nu)$ is the Gamma function. More generally, it can be shown that $\partial^n\sigma_{xy}/\partial H^n\propto L_{\mathrm{eff}}^{n/\nu}$. These formulas can be used in experiments for determining the critical exponent $\nu$.

\begin{figure}[t]
\begin{center}
\includegraphics[width=3.3in,angle=0]{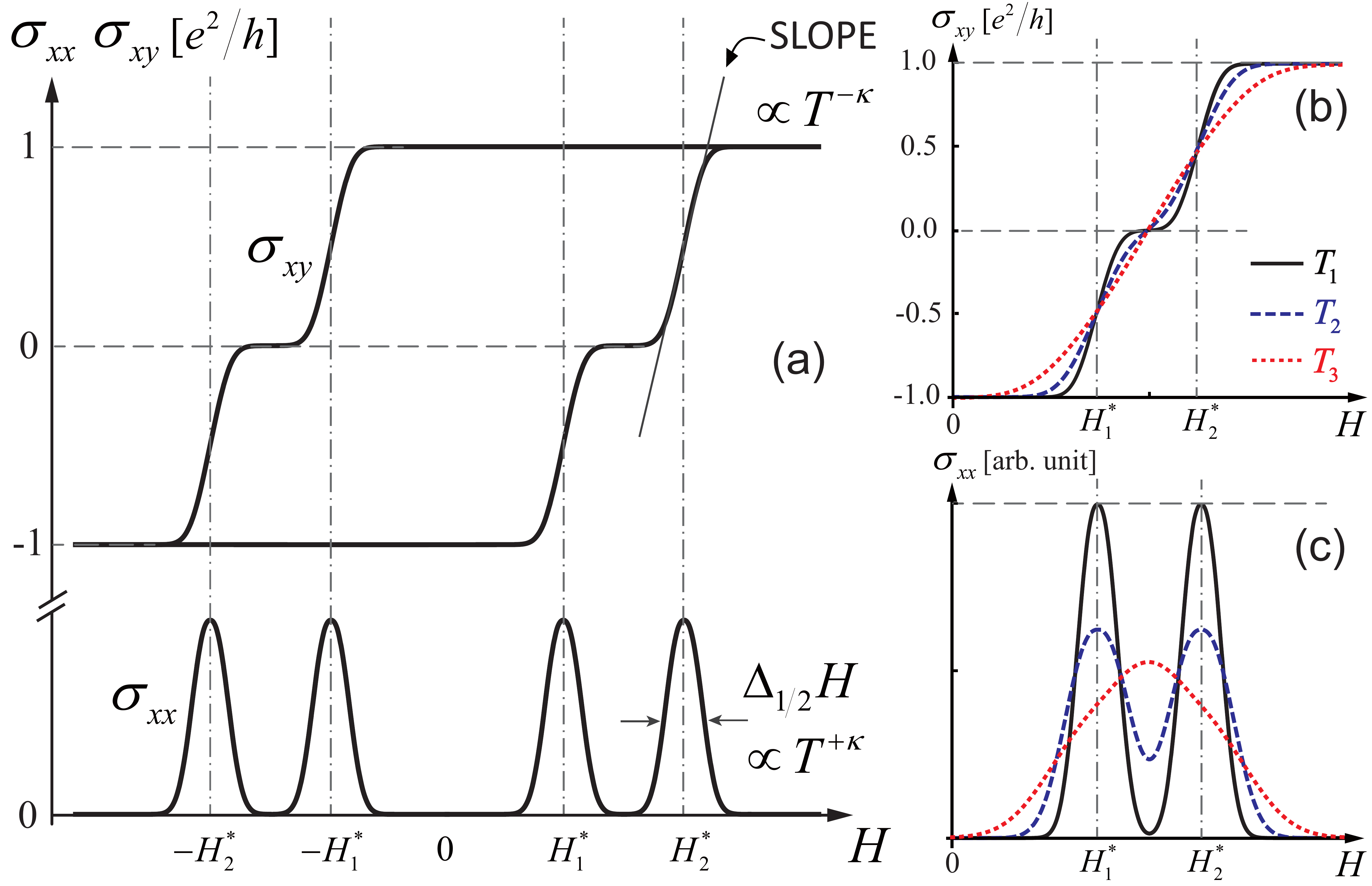}
\end{center}
\caption{(color online) Magnetic field dependence of $\sigma_{xy}$ and $\sigma_{xx}$. (a) Sketch of $\sigma_{xy}$ and $\sigma_{xx}$ as a function of applied magnetic field $H$. The zero quantized plateau appears at the hysteresis loop of $\sigma_{xy}$, while $\sigma_{xx}$ shows two peaks around the coercive field. (b) Sketch of $\sigma_{xy}$ vs. $H$ at three different $T$ with $T_1<T_2<T_3$. (c) The corresponding $\sigma_{xx}$ vs. $H$. Reproduced with permission from~\cite{wang2014a}.}
\label{fig11}
\end{figure}

In the real system, the effective system size is given by $L_{\mathrm{eff}}=\min\{L, a T^{-p/2}\}$, where $L$ is the system size, $T$ is the temperature, $p>0$ is another critical exponent, and $a$ is a constant depending on the system. For sufficiently large system sizes, the maximum slope in the $\sigma_{xy}$ curve then behaves as
\begin{equation}
(\partial\sigma_{xy}/\partial H)_{\max}\propto T^{-\kappa},
\end{equation}
where $\kappa=p/2\nu$. Similarly, the width of the $\sigma_{xx}$ peak scales as
\begin{equation}
\Delta_{1/2}H\propto T^{\kappa}.
\end{equation}

The plateau transition at $\Delta=M$ has the same critical behaviors. In the QAH effect experiment, the exchange field $\Delta$ varies rapidly with respect to the magnetic field $H$ near coercivity, and therefore it is not easy to resolve the two transitions separately. However, with a sufficiently large system size and at low enough temperatures, a zero plateau should be observed in the plateau transition of $\sigma_{xy}$, and $\sigma_{xx}$ should exhibit two peaks, as is shown in Fig.~11. Remarkably, the zero plateau kink in $\sigma_{xy}$ has been observed in recent experiments~\cite{kou2015,fengy2015}.

\subsection{Thickness dependence of the QAH effect}

As discussed in Section~\ref{sec3}, as long as the FM exchange energy $\Delta$ is larger than the hybridization gap $M$ between top and bottom surface states in magnetic TIs, the system will be always in the QAH phase. However, for realistic magnetic TI materials grown on dielectric substrate with different thickness, such condition is not always satisfied.

Clearly, the hybridization gap $M$ and inversion asymmetry $V$ are thickness dependent in magnetic TIs. The thickness of magnetic TI is denoted by the number of QLs ($n$). Qualitatively, $M$ decreases as $n$ increases, while $V$ increases as $n$ increases for such inversion asymmetry can originate from the band bending induced by the substrate. Take (Bi,Sb)$_2$Te$_3$ grown on SrTiO$_3$ (111) substrate for example, $M(n>4)=0$ and $V(n<5)=0$. Therefore, when the thickness is very thin ($n$ very small), $V=0$, $M$ is large, the bulk van Vleck spin susceptibility is greatly reduced due to large hybridization gap, the system would have smaller FM order and even cannot develop FM order, in this case, $M$ will dominate over $\Delta$, the system is in the NI phase. When the thickness is very thick ($n$ very large), $M=0$, $V$ is large and may be dominate over $\Delta$, the system is in the Metal phase [Fig.~2(b)], however, such Metal phase can be tuned into the QAH phase by reducing $V$ through dual gates (top gate and back gate).

In short, when the thickness of magnetic TIs $n\leq n_1$ is small where $M\neq0$ and $|\Delta|<\sqrt{M^2+V^2}$, the system is in the NI phase; when $n\geq n_2$ is large where $M=0$ and $|V|\geq|\Delta|$, the system is in the Metal phase; when $n_1<n<n_2$ where $|\Delta|>\sqrt{M^2+V^2}$, the system is in the QAH phase. However, the exact critical thickness $n_1$, $n_2$ is very much dependent on the
microscopic details of the TI and substrate materials. The thickness dependence of the QAH effect in magnetic TIs predicted here is generic.

\section{Recent experimental progress}
\label{sec7}

Most recently, the QAH effect has been observed by another two experimental groups in the thin films of magnetic TI Cr$_x$(Bi$_{1-y}$Sb$_y$)$_{2-x}$Te$_3$~\cite{checkelsky2014,kou2014}. In the experiment of Tokura's group~\cite{checkelsky2014}, thin films of Cr$_x$(Bi$_{0.2}$Sb$_{0.8}$)$_{2-x}$Te$_3$ are grown on semi-insulating InP(111) substrates using molecular beam epitaxy, which has its Dirac point of the surface states isolated within the 3D bulk band gap. For thin films with a thickness $d=8$~nm and $x=0.22$, the QAH effect is observed at temperatures as low as $50$~mK. In another experiment of Wang's group~\cite{kou2014}, thin films of (Cr$_{0.12}$Bi$_{0.26}$Sb$_{0.62}$)$_2$Te$_3$ are grown by molecular beam epitaxy, and the QAH effect has been observed in both the $10$ QLs thin films and the $6$ QLs thin films at a temperature $T=85$~mK. In both experiments, the Hall resistance $\rho_{xy}$ reaches the quantized value $h/e^2$, while a nonzero residual longitudinal resistance $\rho_{xx}$ remains, which can be explained by the existence of nonchiral edge states as is shown in Section~\ref{transport}.

The experiment of Tokura's group further studied the quantum criticality behaviour near the QAH phase transition, and show evidences that it can be described by the renormalization group theory for the integer QH states. They measured the longitudinal conductance $\sigma_{xx}$ and the Hall conductance $\sigma_{xy}$ at various temperatures $T$ and gate voltages $V_T$. On decreasing the temperature $T$, the renormalization group flow line of $(\sigma_{xy}(V_T),\sigma_{xx}(V_T))$ for a fixed gate voltage $V_T$ can be plotted. These flow lines are in a good agreement with those calculated from the renormalization group theory for the integer QH ground states at fixed Fermi levels, which indicates that the QAH effect and the integer QH effect may have the same critical behaviors. This conclusion is strongly supported by the theoretical analysis presented in Section~\ref{scaling}.

In the experiment of Wang's group, the non-local transport properties of the QAH system are studied. Based on the standard Hall bar shown in Fig.~8(a), they measured the non-local resistance $R_{12,45}$ and $R_{26,35}$, where $R_{ij,kl}$ is defined as the voltage between leads $k$ and $l$ per unit current applied between leads $i$ and $j$. Both of them are non-zero. By flipping the direction of magnetization in the system, they showed that the magnitude of the non-local resistances is changed, which proves the theoretical prediction in Eq. (\ref{R1245}). These observations provided us with strong evidences for the coexistence of nonchiral edge states with the chiral edge state in the QAH system.

\section{Outlook}
\label{sec8}

We have reviewed the theoretical prediction of the QAH in magnetic TIs and discussed recent
theoretical and experimental works in this field. The actual experimental realization of the QAH effect opens up the opportunity
for investigations of many other novel quantum phenomena~\cite{qi2011}.
The nonvanishing zero field longitudinal resistance and the requirement of extremely low temperature for quantization of the anomalous Hall resistance reveal possible delocalized or weakly localized dissipative channels in the system. More measurements are need to confirm the coexistence of chiral and quasi-helical edge states. An intermediate plateau with zero Hall conductance could occur at the coercive field in the QAH plateau transition, and needs further experimental confirmation. The edge channel could be used as a dissipationless spin-filtering path for spintronic devices, and the search for new materials with multiple chiral edge channels would facilitate the device research. Moreover, superconducting proximity with the QAH state would give rise to the chiral topological superconductor~\cite{qi2011}, which supports the Majorana zero mode at the vortex core. With further materials breakthrough, the QAH effect may eventually lead to topological electronic devices of low power consumption.

\begin{acknowledgments}
We thank X L Qi, T L Hughes, C X Liu, H Zhang, X Dai, Z Fang, K He, Y Wang, K L Wang and Q K Xue for collaboration.
This work is supported by the US Department of Energy, Office of Basic Energy Sciences, Division of Materials Sciences and Engineering, under Contract No.~DE-AC02-76SF00515 and the Defense Advanced Research Projects Agency Microsystems Technology Office, MesoDynamic Architecture Program (MESO) through the Contract No.~N66001-11-1-4105, and in part by FAME, one of six centers of STARnet, a Semiconductor Research Corporation program sponsored by MARCO and DARPA.
\end{acknowledgments}

\emph{Note added.} Recently, Kou {\it et al}.~\cite{kou2015} and Feng {\it et al}.~\cite{fengy2015} independently observed the zero Hall plateau state in a QAH insulator when the magnetization reverses, consistent with the theoretical prediction~\cite{wang2014a}.

\end{document}